\documentclass[showpacs,eqsecnum,twocolumn]{revtex4}%preprint,
\usepackage{xspace,psfig,epsfig,amsfonts,amssymb}

\newcommand{\beq}{\begin{equation}}
\newcommand{\eeq}{\end{equation}}
\newcommand{\bqa}{\begin{eqnarray}}
\newcommand{\eqa}{\end{eqnarray}}
\newcommand{\nn}{\nonumber}
\newcommand{\nl}[1]{\nn \\ && {#1}\,}
\newcommand{\erf}[1]{Eq.~(\ref{#1})}

\newcommand{\rt}[1]{\sqrt{#1}\,}
\newcommand{\smallfrac}[2]{\mbox{$\frac{#1}{#2}$}}
\newcommand{\half}{\smallfrac{1}{2}}
\newcommand{\bra}[1]{\langle{#1}|}
\newcommand{\ket}[1]{|{#1}\rangle}

\newcommand{\ito}{It\^o\xspace}

\newcommand{\cu}[1]{\left\{ {#1} \right\}}
\newcommand{\ro}[1]{\left( {#1} \right)}
\newcommand{\an}[1]{\left\langle{#1}\right\rangle}
\newcommand{\implies}{\Longrightarrow}
\newcommand{\raro}{\rightarrow}
\newcommand{\col}{0.45\textwidth}
\newcommand{\erfs}[2]{Eqs.~(\ref{#1})--(\ref{#2})}
\newcommand{\Figs}[2]{Figs.~(\ref{#1})--(\ref{#2})}
\newcommand{\gamr}{\gamma_{{\rm r}}}
\newcommand{\dtime}{\tau_{{\rm dd}}}
\newcommand{\dwj}{d{\cal W}_{{\rm J}}(t)}
\newcommand{\dwjp}{dW_{{\rm J}}(t)}

\begin{document}

\title{Quantum Trajectories for Realistic Photodetection II:
Application and Analysis}

\author{P. Warszawski}
\affiliation{Centre	for	Quantum	Dynamics, School of	Science,
Griffith University,
Brisbane 4111, Australia.}
\author{H. M. Wiseman}
\email{H.Wiseman@gu.edu.au}
\affiliation{Centre	for	Quantum	Dynamics, School of	Science,
Griffith University,
Brisbane 4111, Australia.}

\begin{abstract}
In the preceding paper [Warszawski and Wiseman] 
we presented a general formalism for
determining the state of a quantum system conditional on the output of
a realistic detector, including effects such as a finite bandwidth and
electronic noise. We applied this theory to two sorts of
photodetectors: avalanche photodiodes and photoreceivers. In this
paper we present simulations of these realistic quantum trajectories
for a cavity QED scenario in order to ascertain how the conditioned state
varies from that obtained with perfect detection.  Large
differences are found, and this is manifest in the average of the
conditional purity.  Simulation also allows us to comprehensively
investigate how the quality of the the photoreceiver depends upon its
physical parameters. In particular, we present evidence that in the limit of
small electronic noise, the photoreceiver quality can be characterized
by an {\em effective} bandwidth, which depends upon the level of
electronic noise and the filter bandwidth. We
establish this result as an appropriate limit for a simpler,
analytically solvable, system. We expect this to be a general
result in other applications of our theory.
\end{abstract}

\maketitle

\section{Introduction}

In the preceding paper \cite{WarWis02a} we gave a full 
description of a method to model the 
evolution of open quantum systems conditional upon detection 
results from {\em realistic} detectors. This is a generalization of 
standard quantum trajectory theory, to take into account 
correlations between the system and classical 
detector states which cannot be observed in practice.  We also showed how 
 it can be applied in 
quantum optics, deriving realistic quantum 
trajectories for conditioning upon photon counting using an 
avalanche photodiode (APD), and homodyne detection using a 
photoreceiver (PR).
The greatest significance of our work is in the field of quantum 
control, where conditional states are the optimal basis for control 
loops.

In this paper we find and study solutions to the realistic 
quantum trajectory equations derived in the preceding paper. 
We consider the evolution of a two-level cavity QED 
system (which for convenience we refer to as a two-level atom), 
conditioned on four different types of detection (using the 
two detectors mentioned above). Some of these solutions have been 
studied in a previous paper by us and Mabuchi \cite{WarWisMab01}, 
but not with the 
thoroughness we apply here. These solutions can only be found 
numerically, because of the nonlinearity of the system.  
We also consider another system, the degenerate parametric oscillator 
below threshold (DPOBT), which 
can be treated analytically for realistic homodyne detection. 

Our study in this paper achieves five important aims. First, it 
shows how the theory developed in the preceding paper 
is implementable in practice. 
Second, it establishes the degree of impact of detection imperfections on the 
purity of the conditional system state. Third, it illustrates how   
realistic quantum trajectories typically differ from standard 
(idealized) quantum 
trajectories, which helps to build intuition about them. 
Fourth, it emphasizes the importance of the experimenter's choice of 
detection scheme in 
determining how well one can follow the conditional evolution of a 
system. Fifth, it allows us to quantitatively investigate emergent 
properties of realistic trajectories, such as the ``effective bandwidth'' 
discussed in the preceding paper.

This paper is organized as follows. We begin in Sec.~II by introducing 
the two-level atom (TLA) that will be the subject of study for most of the 
remainder of the paper. We also consider standard (idealized) quantum 
trajectories for this system under the four different measurement 
schemes we consider. In Sec.~III we discuss the idea of levels of 
conditioning. These arise from different levels of knowledge a 
hypothetical observer has about the dynamics of the detector. They 
help one to understand how realistic trajectories differ from idealized 
trajectories. In Sec.~IV we analyze the stochastic dynamics of the 
system under two realistic detection schemes (direct 
and adaptive) using the APD. In Sec.~V we do likewise for the two homodyne 
detection schemes using the PR. In Sec.~VI we return to the question 
of effective bandwidth for the PR, discussed in Sec.~IV~B of 
the preceding paper. 
We verify the formula suggested in the preceding paper, numerically for the TLA 
and analytically for the DPOBT. Sec.~VII contains a discussion of the 
numerical techniques used in our simulations, and Sec.~VIII concludes.

\section{The Two-Level System}

\subsection{The system}

A driven ($\Omega$) and damped ($\gamma$) two-level atom (TLA)
is the smallest (in Hilbert space)
quantum optical system. 
However, it is a nonlinear system and 
has surprisingly rich dynamics. These are especially evident when
one considers conditioning on measurements of its fluorescence
 \cite{DalCasMol92,wtime,WisMil93c,WisToo99}. For this reason, we use it as
a test
 system for investigating realistic photodetection. 
 
 In reality it
 would be almost impossible to collect and detect all, or even most,
 of the fluorescence of an atom in free space. It is therefore better
 to envisage our two-level quantum system as a cavity-QED system
 consisting of a single two-level atom strongly coupled ($g$) to an optical 
cavity
 which is even more strongly damped ($\kappa$). In the limit $\kappa
 \gg g,\Omega \gg \gamma$, the light emitted by the atom into the cavity
 damps through the cavity end-mirror to form an output with the same
 temporal properties as the free-space fluorescence. If $g^{2}/\kappa \gg
 \gamma$, the free-space damping may be ignored and the effective
 damping rate of the two-level system is $\Gamma = g^{2}/\kappa$. We
 have this in mind when choosing parameters for our simulations.
 The advantage
 of this scheme is of course that the cavity output beam is readily
 measurable, by photon counting or homodyne detection.

Let us denote the ground and excited states for the TLA
by $\ket{g}$ and $\ket{e}$.  Defining the Pauli matrices in the usual
manner,  the state matrix for the TLA can be written as
\beq
\rho = \half\left(I+x\sigma_{x}+y\sigma_{y}+z\sigma_{z}\right)\label{rhoXYZ},
\eeq
where $(x,y,z)$ is the Bloch vector which is confined to a unit-sphere.
The purity of the TLA can be defined as
\beq \label{puritydef}
{\rm purity}={\rm Tr}[\rho^{2}]=\half\left(1+x^{2}+y^{2}+z^{2}\right).
\eeq
It has an upper limit of $1$ (a pure state)
and lower limit of $\half$ (a completely mixed state).

The unconditioned ME for the driven and damped TLA
in the interaction picture  is
\beq
\dot{\rho}=-i\frac{\Omega}{2}\left[\sigma_{x},\rho\right]+ {\cal
D}[c]\rho \equiv  {\cal L}\rho.
\eeq
Here $c = \sqrt{\Gamma}\sigma$, where $\sigma$ is the TLA
lowering operator.  The steady state
solution of this ME is
\beq
\rho=\frac{\Omega^{2}+\Omega\Gamma\sigma_{y}+\Gamma^{2}
\left(1-\sigma_{z}\right)/2}
{2\Omega^{2}+\Gamma^{2}},
\label{MESS2}
\eeq
which has a purity of
\beq \label{pME}
p_{{\rm ME}}=1-2\left(\frac{\Omega^{2}}{2\Omega^{2}+\Gamma^{2}}\right)^{2},
\eeq
which goes to $1$ for small $\Omega$ and to $\half$ for large $\Omega$.

\subsection{Perfect Direct Detection}

\begin{figure}
\includegraphics[width=\col]{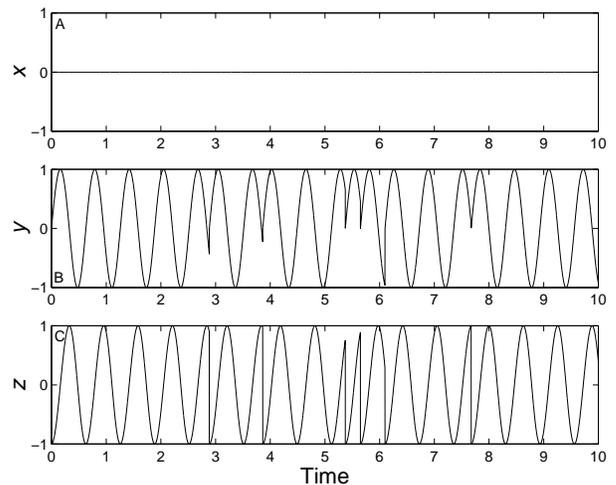}
\vspace{0.2cm}
\caption{A typical trajectory for direct detection for
$\Omega=10\Gamma$.  The two quadratures of the TLA are shown ($x$ in
plot (A) and $y$ in plot (B)) as is the inversion, $z$, of the TLA (in
plot (C).    Time is
measured in units of $\Gamma^{-1}$.  This is true for the
remainder of this paper, unless otherwise stated. }
\protect\label{DDTrajIntroFig}
\end{figure}

Let us now consider continuous, perfect measurement of the TLA.  Firstly
consider counting photons in the emitted field. 
 In the preceding paper we stated the relevant
stochastic master equation (SME) in Sec.~II~C. 
With efficiency $\eta=1$ it is
\bqa
d\rho &=&-\,dt{\cal
H}[iH+\half  c^{\dag}c+\mu^{*}c+\half|\mu|^{2}]\rho 
\nl{+}dN(t){\cal G}[(c+\mu)]\rho. 
\label{rhoNDD}
\eqa
Here ${\rm E}[dN(t)]={\rm Tr}[(c^{\dag}+\mu^{*})(c+\mu)\rho]dt$, 
and $\mu$ is the (necessarily small) 
local oscillator (LO) amplitude. See the 
preceding paper for definitions of the superoperator symbols ${\cal 
H}$ and ${\cal G}$.
A typical quantum trajectory for direct detection (for which $\mu=0$) 
 is shown in
Fig.~\ref{DDTrajIntroFig}, where plots of $x,y,z$ specify the system
state.  

The $x$ quadrature is zero since the Hamiltonian is
proportional to this quadrature.  For this reason we can envisage the
TLA moving in the $y$--$z$ plane of the Bloch sphere.  Since
$\Omega\gg\Gamma$ for this trajectory, the $y$ value of the state
oscillates between $-1$ and $1$ between jumps.  The coherent driving
also causes the inversion of the
TLA, $z$, to oscillate between $-1$ (ground state) and $1$ (excited 
state). Jumps (photon detections) take $z$ to $-1$ 
and $y$ to $0$ as expected.  The conditional purity is of course
unity at
all times.  The
average photon flux entering the detector 
can be calculated from \erf{MESS2} to be
\beq
f_{{\rm direct}} = \frac{\Gamma\Omega^{2}}{2\Omega^{2}+\Gamma^{2}}.
\label{FluxDirect}
\eeq
This saturates at $\Gamma/2$ for $\Omega\gg\Gamma$, as is consistent 
with the number of jumps seen in Fig.~\ref{DDTrajIntroFig}.

\subsection{Perfect Adaptive Detection}
\label{ADDescript}
\begin{figure}
\includegraphics[width=\col]{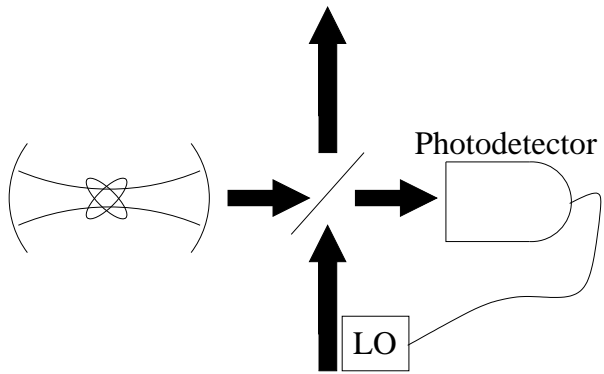}
\vspace{0.2cm}
\caption{Adaptive detection, with the low intensity LO
amplitude controlled by the output of the photodetector.  See text
for further details.}
\protect\label{PDDiagMTAdapt2}
\end{figure}

A quite different sort of conditional dynamics are revealed using 
the adaptive detection scheme proposed in \cite{WisToo99}.
In this scheme the output field from the TLA is combined with a weak
LO at a low reflectivity beam splitter, before being subjected to
direct detection (see Fig.~\ref{PDDiagMTAdapt2}). The local 
oscillator amplitude $\mu$ in \erf{rhoNDD} 
is chosen to be $\pm \sqrt{\Gamma}/2$. 
Two values are present here because 
the LO amplitude is flipped following each
detection. This is what makes the scheme adaptive: the measurement 
being made at any particular time 
depends on the past results.

\begin{figure}
\includegraphics[width=\col]{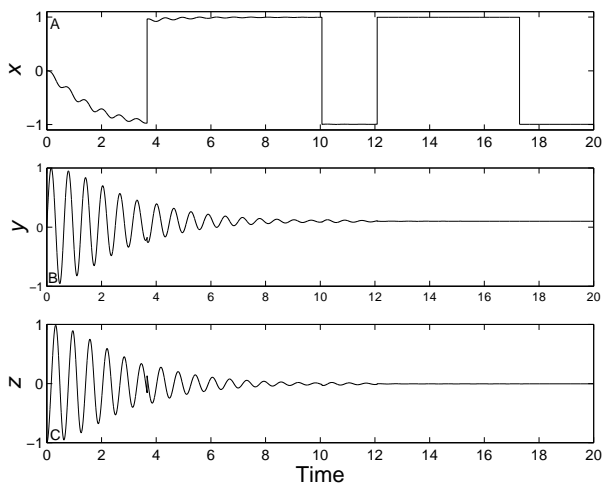}
\vspace{0.2cm}
\caption{A typical trajectory for adaptive detection 
with $\Omega=10\Gamma$.  Two state jumping is clearly visible.}
\protect\label{ADTrajIntro}
\end{figure}

 With this $\mu=\pm\sqrt{\Gamma}/2$ and the $\Omega\sigma_{x}/2$
Hamiltonian,
the behaviour of the TLA is very simple.  After transients have
passed,
the state just jumps between two fixed states that are close to $x$
eigenstates
(for $\Omega\gg\Gamma$).  These two states are actually the stable 
eigenstates of
the two ($\pm$) no-detection measurement operators 
(see, for example, Ref.~\cite{Wis96a}) 
When a detection on the combined field occurs, the TLA jumps into
the stable  eigenstate of the other no jump operator.  A typical
trajectory for adaptive detection is shown in
Fig.~\ref{ADTrajIntro} for $\Omega=10\Gamma$.
The two-state jumping is clearly evident.  Note that $y$ and $z$ take
on the same values in both of the two stable states.

The actual photon flux entering the photodetector for perfect adaptive
detection
can be calculated using the eigenstates of the no-jump operator,
which are given in Ref.~\cite{WisToo99}.  In either eigenstate it is
\beq
f_{{\rm adapt.}} = \frac{\Gamma}{4}.
\label{FluxAdapt}
\eeq
Again, this is consistent with the number of jumps in 
Fig.~\ref{ADTrajIntro} (note the longer time interval displayed).

\subsection{Perfect Homodyne Detection}
\label{HomIntro}
In homodyne detection the LO is so strong that 
individual photons cannot be resolved, and the measurement result is 
a current with white noise $\xi(t)$. For perfect homodyne detection 
the quantum trajectory for the system is generated by the SME 
\beq
d\rho_{I}=dt\left\{{\cal
L}+\xi(t){\cal H}[e^{-i\Phi}c]\right\}\rho,
 \label{rhoI}
 \eeq
which results from the SME in Sec.~II~C~2 of
 the preceding paper with $\eta=1$. 
Here $\Phi$ is the phase of the LO.

\begin{figure}
\includegraphics[width=\col]{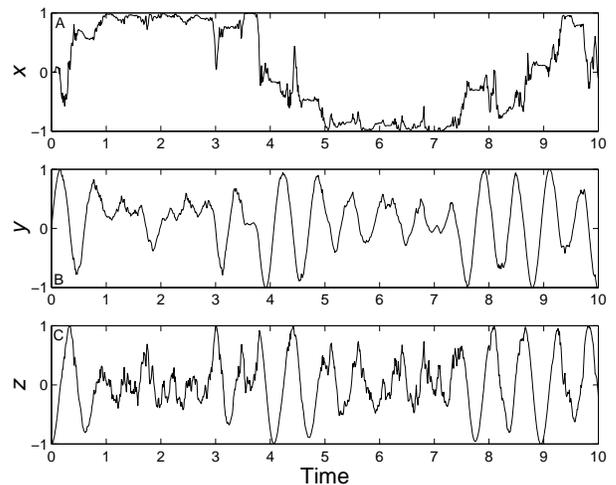}
\vspace{0.2cm}
\caption{A typical trajectory for homodyne $x$ detection with
$\Omega=10\Gamma$.}
\protect\label{HomXTrajIntro}
\end{figure}

Homodyne $x$ detection ($\Phi=0$) corresponds to an `unsharp'
measurement of the $x$ quadrature.  With a $\sigma_{x}$ Hamiltonian,
only the measurement causes $x$ to become non-zero.  Thus, $x$ tends to
be projected into an $x$ eigenstate, with motion between the two
eigenstates on the time scale of $\Gamma$.  This behaviour was first 
noted in Ref.~\cite{WisMil93c}. 
A typical trajectory for $\Omega=10\Gamma$ is shown in  
Fig.~\ref{HomXTrajIntro}. The $\Omega$ oscillations of $y$ and $z$ are 
non-maximal, and are noisy due to the 
white noise in \erf{rhoI}.

\begin{figure}
\includegraphics[width=\col]{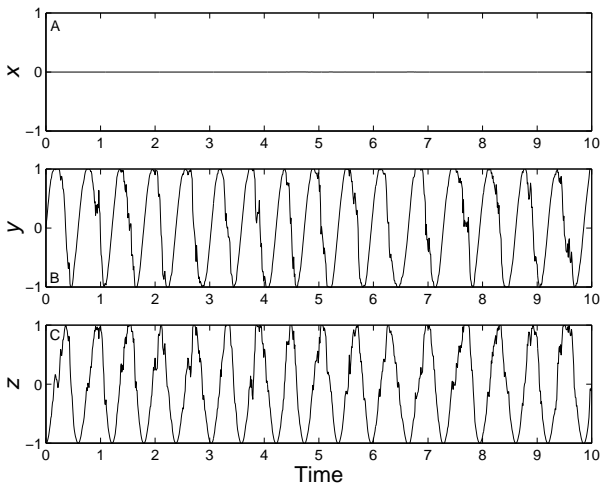}
\vspace{0.2cm}
\caption{A typical trajectory for homodyne $y$ detection with
$\Omega=10\Gamma$.}
\protect\label{HomYTrajIntro}
\end{figure}

In homodyne $y$ detection ($\Phi=-\pi/2$), the $y$ quadrature is 
measured, which pushes
the TLA state closer to the $y$ eigenstates. Unlike for the $x$
measurement though, the $\Omega$ dynamics quickly rotate $y$ away from the
eigenstate. as seen in Fig.~\ref{HomYTrajIntro}.
As there is no measurement or driving of the $x$ quadrature, it 
remains strictly zero, so the oscillations of $y$ and $z$ are maximal.

\section{Levels of Conditioning}

To explicitly show how realistic detection conditions the state of
the TLA differently from perfect detection, it is useful to 
re-introduce the
hypothetical observer of the preceding paper (here known as
 the `perfect' observer). This observer 
 is able to perfectly monitor the system output as it is absorbed 
 by the realistic detector.  A further observer is introduced who is able to access only the microscopic states 
of the detector. 
In the case of the APD this corresponds to knowledge of whether the
state is $0$, $1$ or $2$. For the PR, this `intermediate' observer
knows the voltage across the capacitor.  Hence this observer
avoids some, but not all, of the detector imperfections.
Since the detector is being treated
classically, the perfect and intermediate 
observers   do not affect the average detector state
because there are no detector superpositions to collapse. Hence they 
are consistent with the final `realistic' observer who has 
access only to the realistic experimental records. In the notation of 
the preceding paper, these records are $d{\cal
N}(t)$ and ${\cal V}(t)$ for the APD and PR, respectively. 

Although the measurement records of the three observers are obviously
different, they can be generated {\em from the same experimental run}.
The observers are monitoring the TLA at the same time, but have
varying levels of access to the pool of data that could optimally be
gained from the behaviour of detector.
This means that correlations between the records exist that will be
revealed in the evolution of the TLA that they prescribe.
Through this analysis an understanding of how the
specific features of the trajectories relate to each detector
imperfection can be obtained.

At a particular time each of the three observers will attach, in
general, a different state matrix, $\rho$, to the TLA.  If $\rho$
represents the state of the TLA in an ontological sense then there is
clearly a paradox --- how could the atom have more than one state at
the one time?  The existence of the three observers is, in the
context of the described experiment, not practical, but it is not
difficult to think of a realisable experiment in which one observer must
determine $\rho$ from a `filtered' set of measurement results, while
another has access to the complete set of results.  Another trivial
example is where different observers have different knowledge of
the initial state of the system.  However, in this latter case the
observers, if they had access to the same measurement results, would
eventually agree on the system state.  That is, for a sufficiently
long measurement record, the state is dependent upon the record only
and not the initial condition.  This has been shown by Doherty and
co-workers \cite{Dohe99} and also applies for inefficient
detection.  

In Ref.~\cite{Dohe99}, the authors take the view that the
state is the observer's best description of the system, taking into account
initial knowledge and the measurement record. That is, $\rho$ 
represents the state of the system in an epistemological sense. 
This resolves the paradox raised in the previous 
paragraph, but does not rule out the possibility that 
 there also exists a state of the system in an ontological 
 sense. In particular, the state that the perfect observer assigns to 
 the system is unique because it is a pure state, and no other pure 
 state can be consistently assigned to the system \cite{BruFinMer02}. 
A state vector is the best possible description of the system 
\cite{CavesProb},  and for this reason it seems
that no harm can be done in saying that it represents the `real'
state of the quantum system. The states assigned by the 
intermediate and realistic observers can only be interpreted as 
states of knowledge. 

Before leaving this discussion, it is worth saying a few words about 
consistency. It was already noted above that although 
the state matrices assigned by the three
observers 
are different, they must be consistent.   That is,
the intermediate observer's state must contain the perfect observer's 
state, and the realistic observer's state must contain the 
intermediate observer's state. When we say the state of observer B
 ($\rho_{\rm B}$) contains the state of observer A ($\rho_{\rm A}$), 
 we can write this as
\beq
\rho_{\rm A} \subseteq \rho_{\rm B},
\eeq
which means 
\beq
\exists \epsilon >0 \;:\;\; 
\rho_{\rm B} - \epsilon\rho_{\rm A} \geq 0,
\eeq
and could be read as ``A knows all that B knows''. 
For example, if the perfect observer knows that the TLA is in the ground
state, then the realistic observer would have to say that the TLA is 
in a mixture of the ground state and some other (not necessarily 
orthogonal) state. Note that this is stronger than 
the condition for two states to be mutually compatible, published 
recently in Ref.~\cite{BruFinMer02}, which is simply that they both 
contain a common pure state.

\section{Conditioned Dynamics for Photon Counting}
\label{introPCount}

In this section we will give quantum trajectories for the TLA with 
output  
detected using an APD. We consider both direct detection and adaptive 
detection using a weak LO. We show the system state 
conditioned on three different sets of measurement results,
corresponding to the three observers discussed above.  

The
perfect observer's record contains the times at which photons arrive at the APD.
The evolution is therefore given by \erf{rhoNDD}, 
with $dN(t)$ known. 

The intermediate observer's record consists of the times at which the
various APD transitions occur. Thus the times at which charged pairs are
created, avalanches occur and the APD resets are known.  When a
photon from the quantum system is detected this observer's state will
jump immediately, since the $0\raro 1$ transition is monitored,
instead of displaying the delayed jump that is a characteristic of
realistic detection.

As a consequence, the intermediate observer has access to results
that allow the detector to be described by only two states.
The evolution of the TLA is the same as that which would be the case
for a device that has a
zero response time and a dead time ($\tau'_{{\rm dd}}$)
equal to the random
response time ($\tau_{{r}}$) plus the explicit dead time ($\tau_{{\rm dd}}$).
The equations that are used to evolve the state matrix are similar to
the realistic quantum trajectories for the APD with response rate 
$\gamr\rightarrow\infty$. These equations comprise 
the superoperator Kushner-Stratonovich
equation (SKSE) for the system, which are given in Sec.~III~C of the preceding
paper, but with $\tau'_{{\rm dd}}$ replacing the
deterministic $\tau_{{\rm dd}}$. For each
avalanche, $\tau'_{{\rm dd}}$ could be chosen by generating a random
number $R$ between $0$ and $1$.  Then, based on Poissonian statistics,
\bqa
\tau'_{{\rm dd}}&=&\tau_{{\rm r}}+\tau_{{\rm dd}}\label{tauRand1}\\
&=&- {\ln (R)}/{\gamma_{{\rm r}}}+\tau_{{\rm dd}}.
\label{tauRand}
\eqa

Other changes to the $\gamr\rightarrow\infty$ equations
include replacing $d{\cal
N}(t)$ by $dN_{{\rm cpc}}(t)$, which equals one for the time intervals
in which a charged pair is created (cpc), and replacing the label of
state $2$ by ${\rm dd}$.  This latter change indicates that the APD is
effectively
dead whenever it is in this state.  The modified equations are given
for clarity
\bqa
d\rho_{0}&=&dt\left({\cal
L}-\gamma_{{\rm dk}}-\eta{\cal J}[c+\mu]+
{\rm E}[dN_{{\rm cpc}}(t)]\right)\rho_{0}
\nn\\
&&-dN_{{\rm cpc}}(t)\rho_{0}
+dN_{{\rm cpc}}(t-\tau'_{{\rm dd}})\rho_{{\rm dd}},
\label{dp0No1Ob2}\\
d\rho_{{\rm dd}}&=&dt{\cal
L}\rho_{{\rm dd}}-dN_{{\rm cpc}}(t-\tau'_{{\rm dd}})\rho_{{\rm dd}}\nn\\
&&+d
N_{{\rm cpc}}(t)\frac{(\eta{\cal
J}[c+\mu]+\gamma_{{\rm dk}})\rho_{0}}{{\rm Tr}[\left(\eta{\cal
J}[c+\mu]+\gamma_{{\rm dk}}\right)\rho_{0}]},
\label{dp2No1Ob2}
\eqa
with ${\rm E}[dN_{{\rm cpc}}(t)]=dt{\rm Tr}[(\eta{\cal J}[c+\mu]+\gamma_{{\rm
dk}})\rho_{0}]$.  The superoperator ${\cal L}$  contains the Hamiltonian
evolution as well
as the coupling to the environment, ${\cal D}[c]$.
The intermediate observer cannot distinguish between charged pairs created
by photon absorption or dark counts and is still subject to
an APD inefficiency.

The realistic observer has access only to the times at which
avalanches occur, $d{\cal N}(t)$ (and is able to infer the resetting times).
This observer uses the full SKSE to evolve his/her state, which we restate here
for the reader's convenience
\bqa
d\rho_{0}&=&dt\left({\cal L}-\gamma_{{\rm dk}}-\eta{\cal
J}[c+\mu]+\gamma_{{\rm r}}{\rm Tr}[\rho_{1}]\right)\rho_{0}
\nl{-}d{\cal N}(t)\rho_{0}+d{\cal N}(t-\tau_{{\rm
dd}})\rho_{2}\label{dp0}\\
d\rho_{1}&=&dt\left[\left({\cal L}-\gamma_{{\rm
r}}+\gamma_{{\rm r}}{\rm Tr}[\rho_{1}]\right)\rho_{1}+
\left(\eta{\cal J}[c+\mu]+\gamma_{{\rm dk}}\right)\rho_{0}\right]
\nl{-}d{\cal N}(t)\rho_{1}\label{dp1}\\
d\rho_{2}&=&dt{\cal L}\rho_{2}+d{\cal
N}\left(t\right)\frac{\rho_{1}}{{\rm Tr}[\rho_{1}]}
-d{\cal N}\left(t-\tau_{{\rm dd}}\right)\rho_{2}. \label{dp2}
\eqa

As mentioned
previously, there will be correlations between the three measurement
records $dN(t), dN_{{\rm cpc}}(t)$ and $d{\cal N}(t)$.
Numerical simulation will incorporate these in a way that
will now be discussed.

At each jump of the perfect trajectory a
 check is made to see whether the detector is in the ready state or not.
It will be in the ready
state if ${\rm Tr}[\rho_{{0}}]=1-{\rm Tr}[\rho_{{\rm dd}}]=1$ for the state of
the intermediate observer.
If it is then a
random number ($R'$) between $0$ and $1$
is chosen to see if the photon is actually detected
($R'<\eta$) or not ($R'>\eta$).
If the detection is registered then the stochastic response time, $\tau_{{\rm
r}}$, is
determined according to Poissonian statistics, in the way indicated in
\erfs{tauRand1}{tauRand} and the
transitions that the intermediate observer ($dN_{{\rm cpc}}$) 
and the realistic observer ($d{\cal N}$)
will register are correlated by
\beq
dN_{{\rm cpc}}(t)=d{\cal N}(t+\tau_{{\rm r}})=dN(t).
\label{uq}
\eeq
>From \erf{uq}, the reader is reminded that an avalanche occurs a
random time
$\tau_{{\rm r}}$ after the detected TLA decay. The resetting of the APD occurs
at the same time for the non-perfect trajectories as these observers are
monitoring (with differing levels of expertise) the same device.
If a perfect trajectory jump occurs and the APD is not in the ready state then
no change to the detector takes place.
A brief discussion of the method of integration of the required
differential equations for realistic detection will be given in
Sec.~\ref{NumericSims}.

The $x$ and $y$ quadratures, inversion ($z$) and purity of the
TLA are shown in plots of the trajectories.  The probability of the
detector being in each of its
states is also indicated.  An uncertainty in the state of the
detector exists in the case of realistic detection where the $0\raro 1$
transition is not monitored.

\subsection{Parameter Values for the APD}

In order that the simulations we perform be relevant it is important
that we choose a quantum system and a detector that are realistic.
The quantum system that we are considering is a TLA that has an
effective damping rate given by $\Gamma=4g^{2}/\kappa$, where $g$ is the
TLA-cavity coupling strength and $\kappa$ is the decay rate of the
cavity.  As a guide to the values of these parameters that are
realistically obtainable we use the experiment of Turchette, Thompson
and Kimble \cite{Rice88}, who have $\Gamma=133{\rm
Ms^{-1}}$.  We choose a value of $300{\rm Ms^{-1}}$, which is of the 
same order.

We must also choose values for $\gamr,\dtime,\gamma_{{\rm dk}}$ and
$\eta$.  Realistic values for APDs used in quantum optics 
laboratories are \cite{MabPriv},  $\gamr =7\Gamma$,
$\dtime=2\Gamma^{-1}$, $\gamma_{{\rm dk}}=5\times 10^{-6}\Gamma$
and $\eta=0.8$.  Note that the dark count is negligible.  With these
parameters it will be seen that realistic quantum
trajectories differ significantly from perfect detection trajectories.

\subsection{Direct Detection}
\label{DDCond}

\begin{figure}
\includegraphics[width=\col]{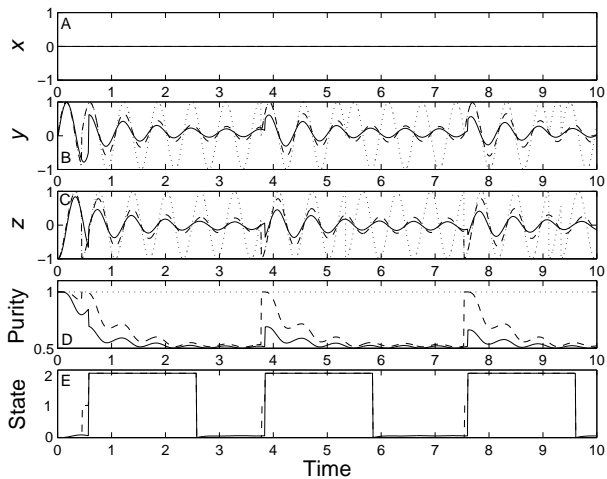}
\vspace{0.2cm}
\caption{A typical trajectory for direct detection.  
The states attached to the TLA are plotted for a perfect observer (dotted
lines), an intermediate observer who can monitor the detector
transitions (dashed line) and a realistic observer who knows only 
when the avalanche reaches a threshold (solid line).  In
plot (E) the occupation probabilities for the various detector states
are shown for the above trajectory, with the same legend applying.
The closeness of the lines to the states $0$ and $1$
indicates the relative probability of occupation. The realistic
parameters used for this photodetection trajectory are 
$\eta=80\%$, and (in terms of the TLA decay
rate $\Gamma=300$M$s^{-1}$) $\gamma_{{\rm r}}=7\Gamma$,
$\tau_{{\rm dd}}=2\Gamma^{-1}$,
$\gamma_{{\rm dk}}=5\times 10^{-6}\Gamma$ and $\Omega=10\Gamma$.}
\protect\label{DirectTraj}
\end{figure}

Trajectories for the three observers are given in
Fig.~\ref{DirectTraj} for direct detection ($\mu=0$).
The dotted line is for the perfect observer
trajectory, the dashed line  for the intermediate observer and the solid line
for the realistic observer.
The perfect trajectory consists of jumps,
 which mean that the TLA must have decayed to  the ground
state, and no-jump evolution, which causes the TLA state to oscillate
coherently between the ground and excited states.

When the atom jumps (here for the sake of simplicity,
we will talk about the perfect trajectory as if that is what the TLA is
actually doing) then the detector will only respond if it is in the
ready state and if the photon absorption leads to the creation of a
charged pair, which occurs a fraction $\eta$ of the time.  At the time
of the first
TLA jump, the detector is almost certainly in the ready state
(dark counts are negligible). From $y$ and $z$ in
Fig.~\ref{DirectTraj} (B) and (C)
we see that
the intermediate observer's state responds immediately, while the realistic
observer has to wait until an avalanche builds up before registering
the detection.  By this time, the TLA would no longer be in the
ground state, so the jump does not take $z$ to $-1$ (or $y$ to $0$).
In fact, the first avalanche actually increases $z$ for the realistic observer,
 contrary to naive expectations.  This occurs because the response rate
($\gamma_{{\rm
 r}}=7$) is smaller than the driving strength ($\Omega=10$) for this
 trajectory.  It is likely that the TLA will have been driven so that it is
 closer to the
 excited state than the ground state in the time taken for an avalanche to
 build up.

After a jump has
occurred, the trajectories separate significantly.  This is because
the detector is `dead' and provides no information about the TLA.  The
evolution during this time for the two non-perfect trajectories 
is via the
unconditioned ME.  The detector can register no more photons until it
is restored to the ready state, a time $\tau_{{\rm dd}}$ after the avalanche
has reached the threshold level.

It can be seen from Fig.~\ref{DirectTraj} (D) that when
a jump occurs, the intermediate
observer's state becomes pure (the impurity introduced by the dark counts is
negligible).  Obviously, this is because it is known that the TLA is the ground
state, which is itself a pure state.
By contrast, the realistic observer's state has a purity of approximately $0.7$
after an avalanche.  Apart from the first avalanche, this represents a
purification.  It arises
because (incomplete) 
information has been gained about the time that the TLA most recently
decayed into the ground state.

As the response rate $\gamma_{{\rm r}}$ of the detector is large compared
to the
 photon flux of the
TLA ($\approx\Gamma/2$ under perfect detection) the
occupation probability of the ready state is close to unity for
times after the detector has been restored to the ready state (see
Fig.~\ref{DirectTraj} (E)).  That is, the transition rate out of state $1$ is
considerably larger than the transition rate into this state.  Another
feature of the
trajectories is that the $x$ quadrature remains at zero
for all time, since the Hamiltonian is proportional to $\sigma_{x}$.
Further details and the physical parameters used for the detector are
included in the figure caption.

The method of simulating the three trajectories for direct detection
was to
firstly unravel the ME for the TLA according to perfect direct 
detection as in \erf{rhoNDD}.  
Thus, a string of jump times at which $dN(t)=1$
 was obtained, along
with the trajectory for the perfect observer.  
The method of correlating the other two
trajectories has already been given in Sec.~\ref{introPCount}.  

%%
 % All that	remains	to be specified	is the evolution
 % that	occurs when	there is no	jump.  As mentioned	earlier,
%when the
 % detector	is dead	(all states	except the ready state)	the
	%evolution is
 % by the ME.  When	the	detector is	in the ready state then	the
 % evolution (unnormalised)	is via \erf{dp0}.
 %%

\subsection{Adaptive Detection}
To show how a realistic APD performs when monitoring
the evolution of a
TLA in a different way, we
consider the adaptive detection scheme described in 
Sec.~\ref{ADDescript}.
In realistic adaptive detection the LO
amplitude will be flipped once the avalanche reaches threshold,
as opposed to the case of perfect detection for which it is flipped at
an earlier time corresponding to when the TLA decays.
Thus, even the perfect trajectory will not
exhibit two-state jumping if we flip the LO at the same delayed time
for all the trajectories.  This is done in order to stay as close as
possible to how the experiment would be performed.

\begin{figure}
\includegraphics[width=\col]{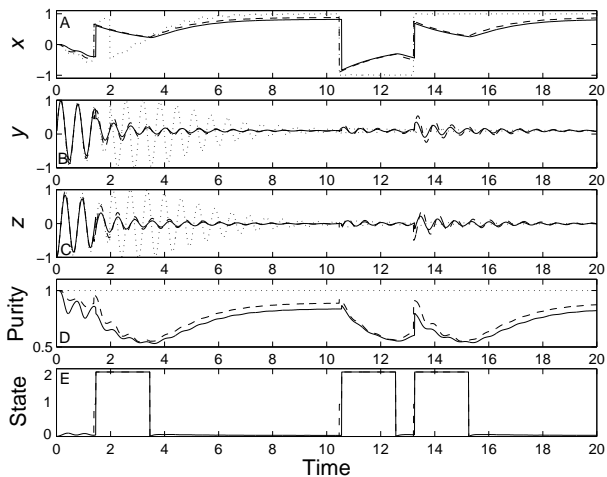}
\vspace{0.2cm}
\caption{A typical trajectory for adaptive detection.  The
simulation parameters and legend are as for Fig.~\ref{DirectTraj}.}
\protect\label{AdaptTraj}
\end{figure}

Trajectories for the same three observers as in direct detection are
shown in Fig.~\ref{AdaptTraj}.  Looking at the $x$ quadrature,
initially the three states head
towards a particular eigenstate of the no-jump operator, then a jump
occurs in the perfect and intermediate trajectories.  A short time after
this the realistic trajectory `avalanches' and closely joins the other
two.  At this time the LO amplitude is flipped.

The TLA (in perfect detection) then closes in on the
second no-jump evolution
eigenstate, while the detector is `dead'.  The non-perfect trajectories
are evolving via the unconditioned ME, away from the eigenstate.
The perfect trajectory then jumps away from the $x=1$ state, implying that a
another photon has been detected in the combined TLA-LO field.
If the LO were
being controlled by the perfect trajectory, then at this point its amplitude
 would be
flipped and the TLA would evolve towards the new eigenstate based on the new
$\mu$.  However, the LO is not `flipped' as this photon goes undetected by the
APD.
The eigenstate of the no-jump evolution, therefore, does not change
and the TLA in the perfect trajectory must evolve back towards it.

A time $\tau_{{\rm dd}}$ after the avalanche,
the detector is ready again and the non-perfect
states head towards that of the perfect trajectory.
The next decay, which happens to be detected by
the APD, takes all
the states to the other side of the Bloch sphere.

The $y$ quadrature and $z$ in (B) and (C) display similar features
when evolving via the ME or the no-jump operator.  This is because the
ME steady state and the operator steady state both have $y$ and $z$ close to
zero, due to the large $\Omega$ being used.  As time
progresses, the three observers' states head towards the steady
states, thus reducing the $y,z$ oscillation amplitude.

One can see from (D) that the purity decreases significantly when the
APD is dead and unconditioned ME evolution is occurring.  When it is
restored to the ready state, the purity
increases as no matter where the TLA is in the Bloch sphere, it will
head towards the no-jump evolution eigenstate.  The average purity is
substantially higher than for direct detection.  This will be
discussed later, when we consider system averages.  Another point is that there
is less difference between the purity of the trajectories of the 
intermediate and realistic observers.  
This is because the state of the TLA is not as
sensitive as direct detection to the time of emission.  The $\Omega$ dynamics
that quickly `swept' the TLA state away from the ground state before the
realistic trajectory jumped are not present in adaptive detection.

The simulations of the adaptive trajectories are more
difficult than those of direct detection, as the perfect
trajectory cannot be run first to determine the decay times.  This is
because the evolution of the TLA depends on the LO amplitude, which
is dependent upon the realistic trajectory.  However, deciding the
jump times of the perfect trajectory from those of the realistic
trajectory is difficult.  The best way to do the simulations
 is to run them
in `parallel' so that all three trajectories are being evolved at the
same time.  The perfect trajectory evolves under the LO determined by
the realistic trajectory, but the avalanches of the realistic trajectory
are determined by the jumps of the perfect trajectory in the same
manner as direct detection.  When the APD avalanches the LO is
inverted for all three trajectories.

\subsection{Average Conditional Purity}

\label{AvePur}

So far we have shown only typical features of realistic detection and 
how they differ from perfect detection. In this section we quantify 
this difference by considering  the steady state average
purity of the conditional TLA state. This is defined as
\beq
p =
\lim_{t\to\infty}{\rm E}\cu{{\rm Tr}[\rho_{\rm c}^{2}(t)]},
\eeq
where the c subscript is included to emphasize that $\rho_{\rm c}$ is 
a conditional state.
Because purity is a nonlinear 
function of $\rho$ (\ref{puritydef}), the steady state ensemble average of the 
conditional purity is not the same as the purity of the steady state ensemble 
average, which is \erf{pME}. The quality factor $p$ 
will be less than one, but  
always greater than $1/2$. The effect of the detector imperfections can
be seen by comparing $p$ for the different measurement schemes
and for a range of driving strengths, $\Omega$.

For small $\Omega$,
even the unconditioned (without measurement) stationary
purity $p_{\rm ME}$ of the TLA approaches unity. To distinguish better
detection schemes in this limit it is useful to define a scaled
purity (between $0$ and $1$)
that measures how much improvement measurement produces:
\beq
{\rm Scaled\;}p=\frac{p-p_{{\rm ME}}}{1-p_{{\rm ME}}}.
\label{ScaledPP}
\eeq
For perfect detection the conditional purity is always 1, 
so the scaled $p$ equals 1 also. For no 
measurement (master equation evolution), the scaled $p$ equals 0.

\begin{figure}
\includegraphics[width=\col]{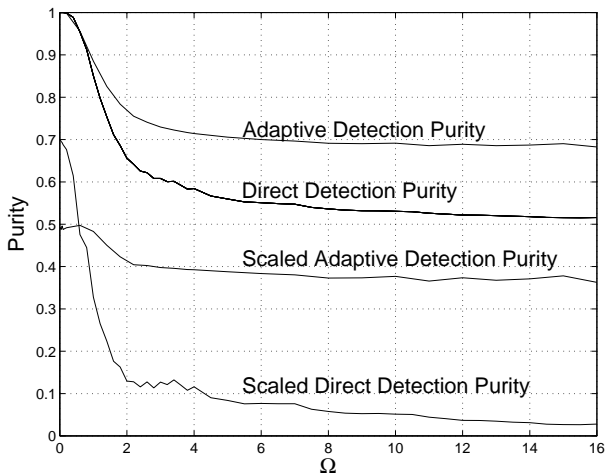}
\vspace{0.2cm}
\caption{The (scaled and unscaled) 
ensemble averaged conditional purity for
realistic direct and adaptive detection as a function of the TLA driving
strength, $\Omega$ (in units of $\Gamma$).
Detector parameters are as for Fig.~\ref{DirectTraj}. In this, and 
subsequent similar plots, the unevenness in the lines is due to, and 
indicates the size of, statistical error.}
\protect\label{PDPurity}
\end{figure}

In Fig.~\ref{PDPurity} we plot the ensemble averaged purity 
(and scaled purity) as a function of the TLA driving strength, $\Omega$,
for realistic detection, direct and adaptive.
The trends in this figure result from two main
imperfections, the dead time and the response time. 

First
consider the input photon flux
in each of the two schemes, given in
\erfs{FluxDirect}{FluxAdapt}.
At low $\Omega$, direct detection produces purer conditioned
states. This is because in this regime
$f_{{\rm direct}}<f_{{\rm adapt.}}$, and direct detection is
therefore less subject to having detections `missed' due to
the dead time  $\tau_{\rm dd}$.  As $f_{\rm direct}$ increases with
$\Omega$, direct detection becomes
worse than adaptive detection as $f_{\rm direct}$  surpasses $f_{\rm 
adapt.}$.  

Note, however, that $f_{\rm direct}$
saturates at $\Gamma/2$,
but the scaled purity for direct detection 
continues to decrease as $\Omega$ increases. 
This is due to the finite detector bandwidth.
Under direct detection, the conditioned state of the TLA Rabi cycles
at rate $\Omega$ (see Fig.~\ref{DirectTraj}).
 The response time $\gamma_{\rm r}^{-1}$ can be thought of
as the uncertainty in the time of photoemission from the TLA, given an
 avalanche. For
$\Omega \sim \gamma_{\rm r}$, the TLA state conditioned on a detection
is `smeared out' by the consequent uncertainty in how far the Rabi cycling
has taken the atom from the ground state since its emission.
By contrast, under
adaptive detection, the scaled purity
asymptotes to a nonzero value because the conditioned dynamics are
governed by $\Gamma$, not $\Omega$ \cite{WisToo99} (see
Fig.~\ref{AdaptTraj}).  Thus the adaptive purity levels off for high 
$\Omega$. From Fig.~\ref{PDPurity},
the difference in the purity of the
trajectories shown for direct and adaptive detection is 
quite large for $\Omega=10$, as expected from the trajectories shown 
in Figs.~\ref{DirectTraj} and \ref{AdaptTraj}.

\section{Conditioned Dynamics For Homodyne Detection}

For the PR, three different levels of `realism' will be
considered, as for the APD.  The first observer once again has the
`perfect' measurement record.  This record is the photon flux incident upon
the p-i-n photodiode.  The evolution of this observer's state is via
\erf{rhoI}.

The intermediate observer has detailed access to the circuit
containing the transimpedance amplifier and is able to determine $I$ 
and $V$ (see Fig.~4 in the 
preceding  paper).  This
means, however, that this observer is still subject to the diode inefficiency.
The current $I$ is given by
\beq
 I=e\sqrt{{\rm P}/\hbar\omega_{0}}\left[\eta\langle e^{-i\Phi}c
+e^{i\Phi}c^{\dag}\rangle +\rt{\eta}\xi'(t)\right],
\label{I}
\eeq
where the parameters are explained in Sec.~IV of the preceding 
paper. The SME for conditioning upon this current is
\beq
d\rho_{I}=dt\left\{{\cal
L}+\rt{\eta}\xi'(t) {\cal H}[e^{-i\Phi}c]\right\}\rho.
\label{rhoIOb2}
\eeq
The white noise $\xi'(t)$ in the above equations 
is related to $\xi(t)$ in \erf{rhoI} by
\beq
\xi'(t)=\rt{\eta}\xi(t)+\rt{1-\eta}\zeta(t).
\eeq
 This is due to the noise arising from two
independent sources: the Poisson statistics of the
LO and the vacuum noise introduced by the
inefficiency of the photodiode.  

To relate this to the realistic observer, we use for the capacitor 
voltage
\beq
\dot{V}' = -\frac{V'}{RC}-\frac{I}{C},
\eeq
or, in terms of the scaled voltage,
\bqa
\dot{v}'&=&-\gamma v'-\sqrt{\frac{\gamma}{N}}\left[\phantom{\sqrt{1-\eta}}
\!\!\!\!\!\!\!\!\!\!\!\!\!\!\!\!\!\!\!
\rt{\eta}\an{e^{-i\Phi}c+ e^{i\Phi}c^{\dag}}
+\sqrt{\eta}\xi(t)\right.\nn\\
&&\left.\phantom{+\sqrt{1-\eta}\zeta(t)}+\sqrt{1-\eta}\zeta(t)\right].
\label{realV}
\eqa
Here, $V'$ represents the unscaled true capacitor voltage.
A prime is used to distinguish it from the argument (dummy variable) 
of the  probability
distribution used by the realistic observer. The expectation value
in \erf{realV} is based on the state matrix obtained through perfect
detection.

The realistic observer will
have access to the output voltage ${\cal V}$ measured in the
laboratory.  The evolution of this observer's state is via the 
SKSE  for $\rho(v)$:
\bqa
d\rho_{{\cal V}}(v)&=&dt\left({\cal
L}+\frac{\gamma}{2N}\frac{\partial^{2}}{\partial
v^{2}}+\gamma\frac{\partial}{\partial v}v \right)\rho(v)
\nl{+}dt\frac{\partial}{\partial
v}\sqrt{\frac{\gamma\eta}{N}}
\left[e^{-i\Phi}c\rho(v)+
e^{i\Phi}\rho(v)c^{\dag}\right]
\nn\\&&+
\rt{\gamma}\dwj
\left(v-\langle v\rangle\right)\rho(v).
\label{dpHom}
\eqa
For simulating the realistic trajectory alone, $\dwj$ would 
be chosen to be an infinitesimal Wiener increment. However, to make it 
consistent with the perfect and intermediate observers, we must use 
the following relation from the preceding paper.
\bqa
\rt{\gamma}\dwj &=& dt\gamma\ro{\rt{\frac{C}{ 4k_{\rm B}T}} {\cal
V}-\an{v}} \\
&=& dt\gamma \rt{\frac{C}{ 4k_{\rm B}T}} \left(V'+\sqrt{4k_{{\rm
B}}TR}\frac{\dwjp}{dt}\right) \nl{-} dt\gamma \an{v} \\
&=& \rt{\gamma}\dwjp +  dt\gamma\ro{ v'-\an{v}}.
\eqa
Here $\dwjp$ is due to the `real' Johnson noise, and is generated as 
an infinitesimal Wiener increment. Thus $v'$ correlates the equation 
for the realistic observer with those for the other two.

Simulation of the three correlated trajectories for homodyne
detection was done in parallel, similarly to adaptive detection.
  This avoids the necessity of storing
the photon flux and $\an{e^{-i\Phi}c+ e^{i\Phi}c^{\dag}}$
from the perfect trajectory at every time step, for later use
in the realistic trajectories.  Once again the reader should see 
Sec.~\ref{NumericSims}
for some of the computer programming details. 

\subsection{Parameter Values for the Photoreceiver}

The same quantum system is monitored as in the case of
photon counting (with $\Gamma=300{\rm Ms^{-1}}$).  For the PR, 
we chose values $\gamma=450{\rm Ms^{-1}}$, $N=0.1$ and $\eta=0.98$. 
These are reasonable values for detectors in quantum optics 
laboratories \cite{MabPriv}. Note that the efficiencies for the 
photodiodes of PRs are
much higher than those of APDs.  This is
because there are various difficulties in ensuring that photon
absorptions lead to avalanches.  Merely sweeping the single
charged-pair out of the depletion region is an easier task.

There is generally a trade-off between 
bandwidth $\gamma$ and the dimensionless 
noise level $N$ \cite{MabPriv}. 
We have chosen a relatively low noise level, and 
consequently a bandwidth below the maximum available. 
This noise level can be related 
to the (more usually quoted) noise equivalent power (NEP) as follows. 
Consider for specificity the PR model\#2007 found in the {\em New
Focus} catalogue \cite{newFoc}.  This model has a 
(NEP) of $\sim
3{\rm pW}/\rt{{\rm Hz}}$ (but a bandwidth of only $790{\rm kHz}$). 
 The NEP can be interpreted as the extra optical 
 power that would need to be injected into the
receiver to simulate noise.  To obtain  $N$ from the NEP, we
must compare it to the LO shot noise that is incident in homodyne
detection.  If the LO has transmitted power of ${\rm P}$ then in
a time interval $\delta t$ the size of the photon number fluctuation
will be $\sim\rt{{\rm P}\delta t/\hbar\omega_{0}}$.  The NEP, on the
other hand, will produce about $({\rm NEP}\hbar\omega_{0})\rt{\delta t}$ 
photons.
Thus, $N={\rm NEP}/\rt{{\rm P}\hbar\omega_{0}}$.
For this PR working close to
saturation, ${\rm P}\approx 0.5{\rm mW}$ \cite{newFoc}, so that
$N\approx 0.1$ for a $\lambda=780{\rm nm}$ LO.

\subsection{Homodyne $x$ Detection}

\begin{figure}
{\bf See attached file 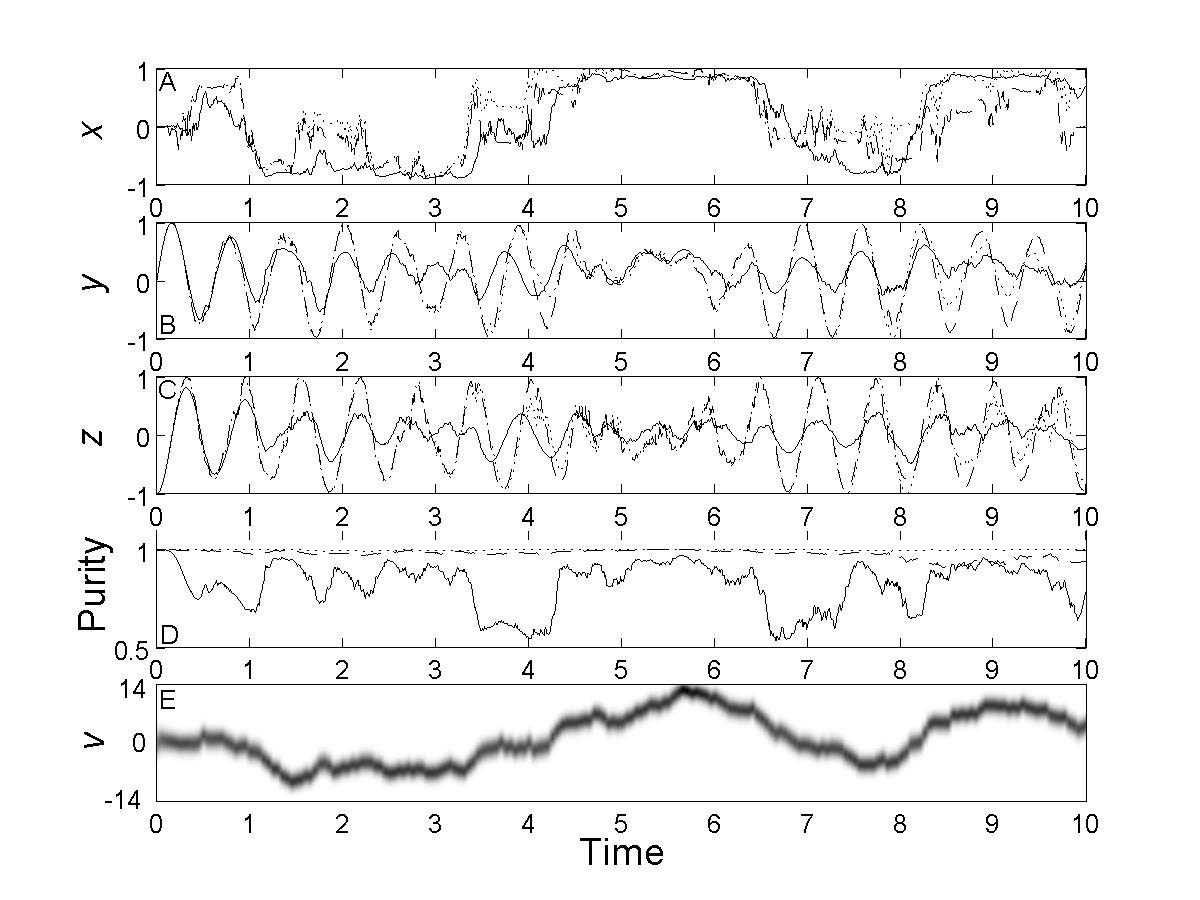}
\vspace{0.2cm}
\caption{A typical trajectory for homodyne $x$ detection.  The
states attached to the TLA are plotted for a perfect observer (dotted
lines), an intermediate observer who knows 
the capacitor voltage $V$ (dashed line) and a
realistic observer who knows only the output voltage ${\cal V}$ (solid line).  
Note that at some times the dotted
and dashed lines are overlapping.  In
plot (E) the occupation probabilities for the scaled capacitor
voltage ($v$)
are given in a grey scale plot. Darker voltages are more likely.
The PR parameters are 
$N=0.1$, $\eta=98\%$, $\gamma=1.5\Gamma$. System parameters are as
for Fig.~\ref{DirectTraj}.}
\protect\label{HomXTraj}
\end{figure}

Trajectories for homodyne detection of the $x$ quadrature ($\Phi=0$)
are given in Fig.~\ref{HomXTraj}.  The line-type allocations are as for
photodetection: dotted is perfect, dashed is the intermediate observer and
solid is the realistic trajectory.  The perfect trajectory has been
described in Sec.~\ref{HomIntro}.

Because the photodiodes of
PRs have a high efficiency (taken to be $\eta=0.98$), the
trajectories for perfect  and intermediate detection 
are very close, especially in $y$ and $z$.  The Johnson noise and response
time 
of the circuit make it more difficult for the realistic observer to
follow the TLA state, although a similar trajectory is still obtained.
The smaller
detail of the perfect trajectory is mostly lost
as the realistic observer has trouble identifying
LO fluctuations that have been filtered and then obscured by Johnson noise.
Another feature is that the realistic trajectory never gets as close as the
perfect trajectory to
being in an $y$ or $z$ eigenstate, reflecting
the mixed nature of the state.  

>From the final subplot (E) of
Fig.~\ref{HomXTraj}, it can be seen that to some extent the
distribution for the scaled capacitor voltage
follows the value of $x$, as one would expect since the current is
proportional to $x$.  Note that the purity dips whenever
there is a large fluctuation
in the $x$ trajectory.  This is indicative of quicker evolution
being more difficult to follow.  

\subsection{Homodyne $y$ Detection}

\begin{figure}
{\bf See attached file 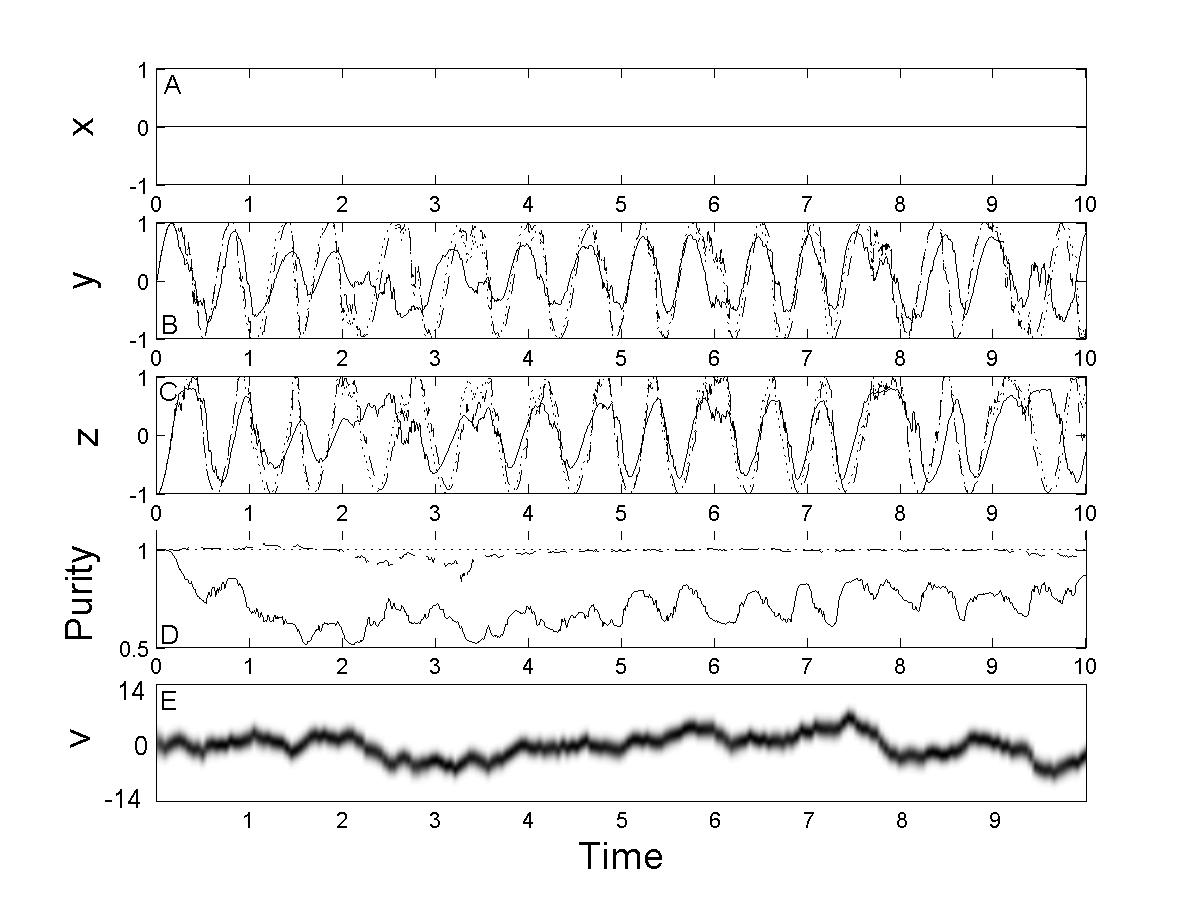}
\vspace{0.2cm}
\caption{A typical trajectory for homodyne $y$ detection.
Legend and parameters are as for Fig.~\ref{HomXTraj}}
\protect\label{HomYTraj}
\end{figure}

The trajectories for homodyne detection of the $y$ quadrature
($\Phi=-\pi/2$) are given in Fig.~\ref{HomYTraj}. Once again
Sec.~\ref{HomIntro} should be referenced for brief comments on the
perfect trajectory.  For homodyne $y$ measurement the trajectory
associated with inefficient detection is close to that of perfect
detection, as it was for the homodyne $x$ measurement.  The realistic trajectory
is a reasonable approximation to the general shape of the perfect
trajectory, although the $y$ quadrature is not being as closely
followed as the $x$ quadrature was for the $x$ measurement.
Note that the amplitude of oscillation of $y$ and $z$ is reduced for
realistic detection.

The distribution for the detector state is influenced by $y$, but
oscillations are
barely visible.  The purity in (D) is lower than the purity for
homodyne $x$ measurement, due the faster evolution of the TLA being more
difficult to
follow.  This will now be discussed in more detail.

\subsection{Average Conditional Purity}

In this section we investigate the long-time ensemble averaged conditional 
purity $p$ as a function of
the driving strength for the PR.  Homodyne measurement of the $x$
and $y$  quadratures of the TLA are contrasted.

\begin{figure}
\includegraphics[width=\col]{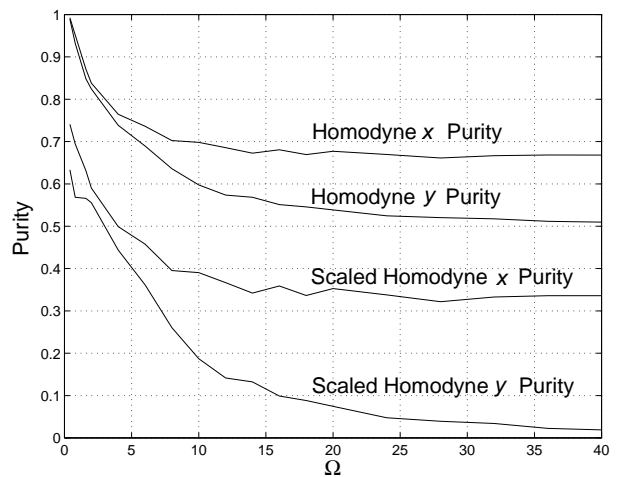}
\vspace{0.2cm}
\caption{The long-time ensemble averaged purity and scaled purity is shown for
realistic homodyne $x$ and $y$ detection as a function of the TLA driving
strength, $\Omega$.  Detector parameters are as for
Fig.~\ref{HomXTraj}.}
\protect\label{PRPurity}
\end{figure}

The results in
Fig.~\ref{PRPurity}
indicate that as $\Omega$ increases,
homodyne measurement of the $y$ quadrature becomes increasingly worse
at following the
evolution of the TLA.  This is due to
the finite bandwidth of the PR in combination with the
conditional homodyne dynamics in the $\Omega \gg
\Gamma$ limit \cite{WisMil93c}.
Homodyne $y$ detection produces a conditional state whose evolution
is dominated  by fast ($\Omega$) Rabi cycling (see
Fig.~\ref{HomYTraj}).
 This is because $x$
is strictly zero, leaving only the $\Omega$ dependent $y$ and $z$
in the expression for $\rho$.

By contrast, homodyne $x$ detection produces mainly slow ($\Gamma$)
dynamics, which can still be tracked by the detector (see Fig.~\ref{HomXTraj}).
  The homodyne
$x$ measurement pushes  $x$ towards the eigenstate ($\pm1$), which
means that $y$ and $z$ must be considerably less than unity as
$x^{2}+y^{2}+z^{2}\leq1$.  The state is, therefore, dependent strongly
on $x$, which is devoid of $\Omega$ oscillation. This explains why 
increasing $\Omega$ beyond a certain point does not cause further loss 
of purity under $x$ detection. These differences are seen more clearly 
in the scaled purity.

\section{Effective Photoreceiver Bandwidth}
\label{effBWAnalysis}

In Sec.~IV~B of the preceding paper we presented the argument that 
for small electronic noise $N \ll 1$, the effective bandwidth $B$ of 
a PR is given not by $\gamma = 1/RC$, but rather by 
\beq
B = \gamma\sqrt\frac{1-N}{N} \simeq \frac{\gamma}{\sqrt{N}}.
\eeq
The meaning of $B$ is that we expect the realistic trajectories from 
the PR to be unable to track system dynamics which have a rate much 
larger than $B$. In this section we investigate this claim by 
studying the realistic quantum trajectories for two very different 
systems. The first is the TLA we have used as our model system so far.
The second is the DPOBT.

\subsection{Two-Level Atom}
\label{EFFBWTLA}
\begin{figure}
\includegraphics[width=\col]{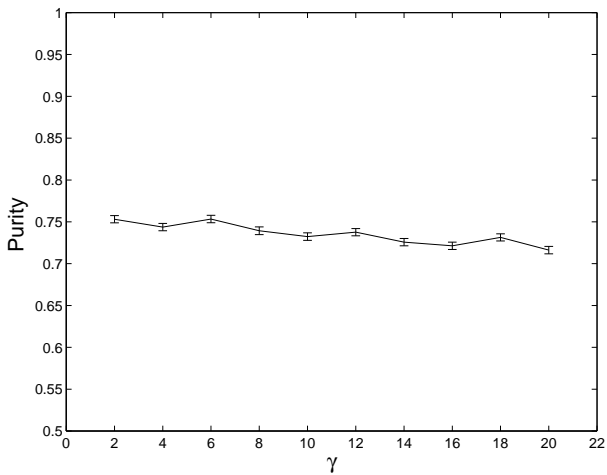}
\vspace{0.2cm}
\caption{The ensemble averaged purity  as a function of $\gamma$ for
realistic homodyne $x$ detection when $\gamma\sqrt{1-N}/\sqrt{N}$ is
kept constant at a value of $20$.  The efficiency is $\eta=98\%$ and
the driving strength of the TLA is $\Omega=30$.  Error bars are
included to make it clear that the slight downward slope is not due to
fluctuations. }
\protect\label{effBWFig}
\end{figure}

To test the prediction of the effective bandwidth
the ensemble purity of the TLA was calculated for a
range of $\gamma$ and $N$, while maintaining the proposed quality 
indicator of the
PR, $\gamma\sqrt{1-N}/\sqrt{N}$, as a constant.
Our theory predicts that the purity will remain constant.  The results
for homodyne $x$ measurement are
contained in Fig.~\ref{effBWFig} and show that the purity is indeed
relatively flat, when it is considered that $N$ is 
 varying by almost two orders of magnitude.

There is nevertheless a slight downward trend in the graph in 
Fig.~\ref{effBWFig}. This is most likely
due to the presence of increased noise as $\gamma$ (and $N$) increases. 
As explained in the preceding paper, our argument 
only   makes sense for small $N$. 
In the limit of $\gamma\rightarrow\infty$ we
are returning to the situation of adding noise only (discussed in the 
preceding paper), 
which is equivalently described by
an inefficiency of $\eta/(1+N)$. Thus even if $B \to \infty$ as 
$\gamma\to\infty$, a finite noise $N$ would still reduce the system 
purity. 

 If
$B$ could be kept constant with $N$ decreasing to
zero then we would expect our prediction to become exact.   
Unfortunately this is a difficult
regime to test numerically as it is actually the ratio of $\gamma/N$
not $\gamma/\sqrt{N}$ that appears in \erf{dpHom}.  With
$\gamma,N\rightarrow 0$, but $B$ constant,
$\gamma/N\rightarrow\infty$.  The time step involved in the
simulation decreases and the number of iterations of \erf{dpHom}
increases, causing the computational intensity to become
prohibitive (see Sec.~\ref{NumericSims}).  

Further support for identifying $B$ with the effective bandwidth is
found in Fig.~\ref{PRPurity}, where purity and scaled purity are plotted
against $\Omega$.  It is expected that approximately half the total
loss of purity will have occurred once $\Omega\sim B$.  
That is, once the frequency of the monitored signal becomes
equal to the effective inverse response time of the receiver.  With
the parameters of Fig.~\ref{PRPurity} we have
$\gamma\sqrt{1-N}/\sqrt{N}=4.5$, which is in approximate agreement
with the value of $\Omega$ such that $p=(p_{{\rm max}}+p_{{\rm
min}})/2$.  Only approximate agreement is expected because of the
complicating issue of noise, as discussed previously.

\subsection{Degenerate Parametric Oscillator}

Although the TLA is a small system, amenable to numerical simulations, 
it is usually not possible to find analytical solutions for its quantum 
trajectories. This is true of 
perfect detection schemes, let alone realistic detection schemes.
We have seen above that even numerically it is difficult to do 
simulations in a regime of theoretical interest, $\gamma, N \to 0$ such 
that $B$ is finite. 
 For this reason, we now turn to
  a simpler quantum system, the 
  degenerate parametric oscillator below threshold.

The DPOBT system consists of a damped single-mode optical cavity 
containing a $\chi^{(2)}$ nonlinear crystal which is pumped by a 
(classically described) laser at twice the cavity frequency. We take 
the intensity damping rate to be unity and the parametric driving 
strength, $\chi$, to be of modulus less than unity. This
leads to squeezing in the $y$ quadrature, $-i(a-a^{\dag})$,
 if $\chi>0$ and squeezing in the $x$
quadrature, $a+a^{\dag}$, if $\chi<0$.  The system  obeys the
following ME
\beq
d\rho =dt\left(-\smallfrac{1}{4}\chi[a^{2}-a^{\dag 2},\rho]+{\cal
D}[a]\rho\right),
\eeq
when no measurement is performed.

We will limit ourselves to considering realistic homodyne measurement of the
$x$ quadrature.  Using \erf{dpHom} with $c=a$, the SKSE is
\bqa
d\rho_{{\cal V}}(v)&=&dt\left(-\smallfrac{1}{4}\chi{\cal H}\left[a^{2}-a^{\dag
2}\right]+{\cal D}[a]\phantom{\frac{\partial^{2}}{\partial
v^{2}}}\right.\nn\\
&&\left.\phantom{+{\cal D}[a]+{\cal D}[a]}+\frac{\gamma}{2N}\frac{\partial^{2}}{\partial
v^{2}}+\gamma\frac{\partial}{\partial v}v \right)\rho(v)\nn\\
&&+\frac{\partial}{\partial
v}\sqrt{\frac{\gamma\eta}{N}}
\left[a\rho(v)+\rho(v)a^{\dag}\right]\nn\\
&&+
\rt{\gamma}\dwj\left(v-\langle v\rangle\right)\rho(v).
\label{dpHomCav}
\eqa

The superoperator can be removed by converting this
into a Kushner-Stratonovich equation 
(KSE) for the probability distribution,
$P(x,v)$,
for $x$ and $v$. This is done
 using the standard procedure for Wigner functions presented
in Ref.~\cite{GarNoise}.  The distribution is defined by
\beq
P(x,v)=\bra{x}\rho(v)\ket{x},
\eeq
with $\ket{x}$ being an eigenstate of the $x$ quadrature operator.
Due to the linear nature of the optical cavity the
measurement of the $x$ quadrature does not disturb the distribution for the $y$
quadrature after initial transients have died away.
Thus, all statistics concerning the $y$ quadrature can be obtained
from the unconditioned ME.  In particular, the variance of the $y$ quadrature,
$\Delta_{y}$, in the steady state of the ME is
\beq
\Delta_{y}=\frac{1}{(1+\chi)^{2}}.
\label{quady}
\eeq

Conversion of \erf{dpHomCav} gives
\bqa
dP_{{\cal V}}(x,v)&=&dt\left[k\frac{\partial}{\partial x}x
+\frac{1}{2}\frac{\partial^{2}}{\partial x^{2}}
+\frac{\gamma}{2N}\frac{\partial^{2}}{\partial
v^{2}}+\right.\nn\\
&&\left.\gamma\frac{\partial}{\partial v}v +
\sqrt{\frac{\gamma\eta}{N}}\frac{\partial}{\partial
v}
\left(x+\frac{\partial}{\partial x}\right) \right]
P(x,v)\nl{+}
\rt{\gamma}\dwj\left(v-\langle
v\rangle\right)P(x,v),
\label{dWHomCav}
\eqa
where, $k=\half(1-\chi)$, 
and the expectation value is found from
\beq
\an{v}=\int dxdv\: vP(x,v).
\eeq
  The damping term has turned into drift and
diffusion in $x$, while the parametric driving has become a drift term. It
should be noted that in this section we are using $x$ to represent the
possible values that the $x$ quadrature can take, rather than being
the mean of the quadrature, which will be denoted by $\an{x}$.

Despite its nonlinear and stochastic nature, a SKE of the form of 
Eq.~(\ref{dWHomCav})  
has analytical long-time solutions, a fact which is at the heart of 
modern engineering control techniques 
\cite{Jac93}. These solutions are Gaussians, and 
a closed set of equations of motion exist for the
conditioned mean vector and covariance matrix for $x$ and $v$. 
  The equation for
the covariance matrix does not depend on the mean vector, 
 and is also deterministic,
thus having a steady state solution. We denote the three elements of 
the covariance matrix
\bqa
\Delta_{x}&=&\an{x^{2}}-\an{x}^{2},\\
\Delta_{v}&=&\an{v^{2}}-\an{v}^{2},\\
\Delta_{xv}&=&\an{xv}-\an{x}\an{v}.
\eqa
It is important to remember the ``\ito correction'' \cite{Gar85} in 
calculating the equation of motion for these quantities. For example,
\beq
d\Delta_{x} = d\an{x^{2}}-2\an{x}d\an{x}-d\an{x}d\an{x}.
\eeq

Using integration by parts and the vanishing of 
$P(x,v)$ at infinity yields
\bqa
d\an{x}&=&-k\an{x}dt+\sqrt{\gamma}d{\cal W}_{{\rm J}}(t)\Delta_{xv}\label{x}\\
d\an{v}&=&-\left(\gamma\an{v}+\sqrt{\frac{\gamma\eta}{N}}\an{x}\right)dt+
\sqrt{\gamma}d{\cal W}_{{\rm J}}(t)\Delta_{v}\label{v}\nn\\
\\
d\Delta_{x}&=&(-2k\Delta_{x}+1-\gamma\Delta_{xv}^{2})dt\label{Vx}\\
d\Delta_{v}&=&\left(\frac{\gamma}{N}-2\sqrt{\frac{\gamma\eta}{N}}\Delta_{xv}
-2\gamma\Delta_{v}-\gamma\Delta_{v}^{2}\right)dt\label{Vv}\\
d\Delta_{xv}&=&-\left[(k+\gamma)\Delta_{xv}+\phantom{
\sqrt{\frac{\gamma\eta}{N}}}\right.\nn\\
&&\left.\phantom{\Delta_{xv}}
\sqrt{\frac{\gamma\eta}{N}}\left(\Delta_{x}-1\right)
+\gamma\Delta_{v}\Delta_{xv}\right]dt,
\label{Vxv}
\eqa
which are examples of Kalman filter equations \cite{KMan}.
In the long time limit the covariances are constant, and 
the whole distribution is just
being shifted with the motion of the mean.

Since the purity of
the cavity mode is defined in terms of the variances of the $x$ and
$y$ quadratures, after initial transients have died
the purity will be constant in time. For Gaussian states with no
correlation between the $x$ and $y$ quadrature, the purity ${\rm
Tr}[\rho^{2}]$ is given by
\cite{HowVac}
\beq
p=\frac{1}{\rt{\Delta_{x}\Delta_{y}}}.
\eeq
There is an analytical steady state solution of \erfs{Vx}{Vxv}, but 
it  is very complex and 
 will not be given here. We are really only interested in the
purity in the limit of small $N$ with $B = \gamma/\sqrt{N}$ fixed.
If $B$ is indeed the effective bandwidth, as we have argued, then in
the limit $N\to 0$ with $B$ fixed, the purity should depend on $B$
only, not $\gamma$ or $N$.

To show this we will examine \erfs{Vx}{Vxv} more closely.  The
first step is to determine how $\Delta_{x},\Delta_{v},\Delta_{xv}$ scale
as $\gamma,N\raro 0$.  Re-writing \erfs{Vx}{Vxv} in terms of $B$ and
$N$, we have from \erf{Vx} at steady state
\beq
\Delta_{x}=\frac{1-B\sqrt{N}\Delta^{2}_{xv}}{2k}.
\eeq
>From the unconditioned ME, $\Delta_{x}$ would be $1/2k$, while we
expect (and have verified numerically) that a finite $B$ will give a finite
variation of the purity away from $p_{{\rm ME}}$.
  For this to be the case it must be true (when  $\gamma,N\raro 0$)
that $\Delta_{x}\lesssim 1/2k$ with 
\beq
B\sqrt{N}\Delta^{2}_{xv}\sim \sqrt{N}\Delta^{2}_{xv} \sim 1
\;\implies\; \Delta_{xv}\sim {N^{-1/4}}.
\label{Vxvapprox}
\eeq
Ignoring the $\gamma$ term in \erf{Vxv} that is small compared to
$k$ gives
\bqa
\Delta_{v} &=& -\left[\frac{k}{B\sqrt{N}}+
\frac{1}{\Delta_{xv}N^{3/4}}\sqrt{\frac{\eta}{B}}\left(\Delta_{x}-1
\right)\right]\nn\\
&\sim& N^{-1/2},
\eqa
where we have used \erf{Vxvapprox}.

New variances of order unity are now defined:
\bqa
\tilde{\Delta}_{v}&=&N^{1/2}\Delta_{v}\\
\tilde{\Delta}_{xv}&=&N^{1/4}\Delta_{xv}.
\eqa
The variance in $x$ is kept the same, as it is of the order unity.
Using these in \erfs{Vx}{Vxv} and ignoring small terms allows these
equations at steady state to be written as
\bqa
-2k\Delta_{x}+1-B\tilde{\Delta}_{xv}^{2}&=&0\label{VxN}\\
B-2\sqrt{B\eta}\tilde{\Delta}_{xv}-B\tilde{\Delta}^{2}_{v}&=&0\label{VvN}\\
-k\tilde{\Delta}_{xv}-\sqrt{B\eta}(\Delta_{x}-1)-B\tilde{\Delta}_{xv}\
\tilde{\Delta}_{v}&=&0\label{VxvN}.
\eqa
Importantly, these equations, which apply in the $\gamma,N\raro 0$
limit, are only dependent upon the detector parameters $B$ and $\eta$.
This means that the $\gamma,N$ dependence of the
purity can be summarized by the one parameter $B=\gamma/\sqrt{N}$. 
This confirms the correctness of our argument.

As \erfs{VxN}{VxvN} are more simple than \erfs{Vx}{Vxv} it would be
useful to solve for $\Delta_{x}$, and hence purity.  This can be
done, using the facts that the variances are positive and that 
$\Delta_{x}\Delta_{y}\ge 1$ to discard non-physical solutions.
 The result for purity  is
\beq
p=\ro{\frac{{2k\eta(1-k)B^{2}}}
{{S\sqrt{\eta
B^{3}}\left(BR+k^{2}\right)-k^{4}-B^{2}k^{2}
\left(1-\frac{\eta}{k}+\frac{2R}{B}\right)}}}^{1/2},
\label{pB}
\eeq
with
\bqa
R&=&\sqrt{k^{2}+\eta(1-2k)}\nn\\
S&=&k\sqrt{\ro{2BR+k^{2}+B^{2}}/{\eta B^{3}}}.
\eqa

We can now consider the two obvious limits of $B\raro 0,\infty$ (with
$\eta=1$ for simplicity).
For $B\raro 0$ the purity is, to lowest order
in $B$,
\beq
p=p_{{\rm
ME}}+\frac{B^{2}}{4}\frac{(1-2k)^{2}(1+k)^{2}}{k^{7/2}(1-k)^{3/2}}.
\eeq
As
$B\raro\infty$ the purity to first order in $1/B$ is
\beq
p=1-\frac{1}{4B}\frac{(1-2k)^{2}}{(1-k)}.
\eeq
That is, for small $B$, the purity increases from the unconditioned purity
quadratically with the effective bandwidth, while for large $B$
it decreases from
unity linearly in $B^{-1}$.

As a final point, it is worth noting that the downward trend
observed in the purity in Fig.~\ref{effBWFig} for the TLA 
is also borne out in the DPOBT.  
That is, for fixed $B=\gamma\sqrt{1-N}/N$, the purity still
has a slight downward trend with $N$.  This was found through numeric investigation of the complete solution for the purity (which was not stated due
its complexity).  However, the numeric investigation also revealed that
as $N\to 0$ the purity curve
maintains the same slope, going to the limits found analytically above.

\section{Numerical Simulations Technique}
\label{NumericSims}
%numlabel

Before concluding, we here comment explicitly upon the method of 
numerical simulation of the SKSE's contained in this paper.
Obtaining realistic trajectories from \erfs{dp0}{dp2} and \erf{dpHom}
is obviously a non-trivial numerical exercise,
even when they do not have to be correlated with the unrealistic 
trajectories of the perfect and intermediate observers.
To set up the problem, the {\em Quantum Optics toolbox for Matlab}
\cite{QOTool} was used.  This allowed easy formulation of the
required quantum (super)operator expressions.

We represented the
supersystem state with a long column vector and a square matrix, ${\bf L}$,
was
used to evolve it.  That is,
\bqa
d\left(\begin{array}{ll}
\rho_{1}\\
\vdots\\
\rho_{s}\\
\vdots
\end{array}\right)=dt{\bf L}\left(\begin{array}{ll}
\rho_{1}\\
\vdots\\
\rho_{s}\\
\vdots
\end{array}\right),
\label{Numerf}
\eqa
where
\beq
\rho_{s}=
\left(\begin{array}{ll}
\rho_{s, e}\\
\rho_{s, ge}\\
\rho_{s,eg}\\
\rho_{s, g}
\end{array}\right).
\eeq
The subscripts $e,g$ represent the ground and excited state of
the TLA.  The integer subscripts $s$ label detector states $s\in
\mathbb{S}$.  For the APD there are only three detector states,
while for the PR the scaled capacitor voltage is discretized on a
grid.  The matrix ${\bf L}$ generates 
all the evolution on the RHS of \erfs{dp0}{dp2}
and \erf{dpHom}.

Because of the stochastic nature of the realistic
trajectories, ${\bf L}$ cannot in general 
be formed in its entirety at the start
of the simulation.  In the case of the APD one can avoid this difficulty by 
using the unnormalized versions of \erfs{dp0}{dp2}.  These were given in 
\cite{WarWisMab01}.  The stochasticity then enters via the comparison of the 
norm of the supersystem state to a random number in order to choose the 
avalanche times.  For the PR, the nonlinear term (the expectation value, 
$\langle v\rangle$) is included in the evolution.  This, in addition 
to the noise, $\dwj$, means that the best that can be 
done is to create the ${\bf L}$ for all but the last term of \erf{dpHom}.
The last term is created every time step after the calculation of the 
expectation value.

The structuring of the numeric solution in this way 
(creating as much as possible of ${\bf L}$ before the iteration process begins)
makes it
very flexible.  For example, a change in the nature of the Hamiltonian
for the TLA could be easily achieved by creating the new
superoperator in the Quantum Optics toolbox and thus obtaining a
new ${\bf L}$.  This is in contrast to the technique of
deriving
$\dot{x}_{s},\dot{y}_{s},\dot{z}_{s}$ equations and then evolving
these numerically, as a new Hamiltonian would require a new derivation.

The obvious disadvantage with ${\bf L}$ is that it will be very
sparse in general.  This is overcome by reducing it to only the
non-zero elements, with the aid of the {\em find} command in {\em
Matlab}.
Once the reduced matrix has been found it is written to a data file
which is used as the input for a {\em C}++ program.

The {\em C}++ program then integrates the matrix equation \ref{Numerf}
 by
looping through all the non-zero ${\bf L}$ elements and making the
appropriate increments.  The simple Euler technique of integration is 
used, by which is meant that the infinitesimal $d$ in \erf{Numerf} is 
replaced by the small but finite $\delta$. 
This method has well known instabilities, but is used by other
workers in the field of quantum optics.  
It has the property of being accurate when
the solutions do not explode, as opposed to more stable methods which
have less spectacular failures that are more difficult to 
detect \cite{EulerCarmichael}.

Some specific details of the PR simulations are as follows.
A stationary grid of $100$ points was used for the capacitor
voltage distribution.  These points were spread $7$ standard
deviations of the initial distribution either side of the initial mean.
The initial distribution was found by solving the
Ornstein-Uhlenbeck equation \cite{Gar85} 
\beq
dP(v) = dt\left(\frac{\gamma}{2N}\frac{\partial^{2}}{\partial
v^{2}}+\gamma\frac{\partial}{\partial v}v \right)P(v).
\eeq
This is derived from the SKSE (\ref{dpHom}) by removing the TLA and 
averaging over the realistic measurement. Ignoring the TLA 
is reasonable as the field from the 
TLA is of the same order as the vacuum field. 
This leads to a Gaussian solution
with a variance of
$\Delta_{v,{\rm u}}=1/2N$, giving a standard deviation of 
 $2.24$ for $N=0.1$.  The u subscript is to indicate
that this is a variance unconditioned on measurement.
From \Figs{HomXTraj}{HomYTraj}, it can be seen that the considered range of
voltages was sufficient.

A moving grid for the voltage distribution was considered but 
 {\em not} used in the end.  This decision 
can be justified by calculating the {\em conditioned} variance of the
voltage distribution in the case of a vacuum input.  Unlike the
calculation of the variance from the Ornstein-Uhlenbeck equation,
the stochastic measurement term 
\beq
\rt{\gamma}\dwj\left(v-\langle
v\rangle\right)P(v)
\eeq
is now included. 
The variance goes to a steady state value, despite the stochasticity.
The derivation of the conditioned variance, $\Delta_{v,{\rm c}}$, is
performed in the same manner as in Sec.~VI~B, and the result is
\beq
\Delta_{v,{\rm c}}=\rt{1+\frac{1}{N}}-1.
\label{VCondA}
\eeq
This is always smaller than the unconditioned variance, $1/2N$.  In
fact, in the limit $N\raro 0$
\beq
\frac{\Delta_{v,{\rm u}}}{\Delta_{v,{\rm c}}}=\frac{1}{2\rt{N}}.
\eeq
In this limit the conditioned variance becomes much smaller than the
unconditioned variance and a moving grid would save much numerical
computation.  This is because the number of grid points
necessary to describe the non-zero probabilities at a particular time
is much less than those required to describe the movement of the distribution
over all time.  However, for $N=0.1$ we have $\Delta_{v,{\rm u}}=2.16
\Delta_{v,{\rm c}}$ and the saving is not very large.  It is worth
noting that a moving grid would not solve all the computational
intensity of $\gamma/N\raro 0$ as the problem of the decreasing
required time step still exists.

For the PR, time steps of $\delta t=1\times 10^{-5}\Gamma^{-1}$
generally proved satisfactory, as did
ensemble sizes of about $1000$ for forming averages.
The time step for the APD can be increased
by about an order of magnitude as there is no white noise.
The ensembles actually took the
form of a collection of samples of the supersystem
state in one trajectory, taken at times separated by $\Gamma^{-1}$.
This is large compared to the
system correlation time \endnote{The smallest non-trivial negative
real part of an eigenvalue of the
Liouvillian is $\Gamma/2$, so samples separated by $\Gamma^{-1}$
should be sufficient.}.  The equivalence of this time averaging to
a many trajectories average has been established by Cresser
\cite{Cresser}.

\section{Conclusion}

In this paper we have applied the theory of realistic quantum 
trajectories for photodetection 
derived in the preceding paper \cite{WarWis02a} to realistic quantum optical 
situations. We investigated two systems. The first was a driven two-level 
system, which could be realized as 
a strongly coupled atom-cavity system, heavily damped though one
cavity mirror. For this system we solved for the trajectories 
numerically. We looked at features of typical stochastic 
trajectories, as well as a figure-of-merit, the ensemble averaged 
conditional purity, for how well the system state is known. We 
considered four different detection schemes: direct  
and adaptive detection using an avalanche photodiode (APD), and $x$ and $y$ 
quadrature homodyne detection using a photoreceiver (PR). The second 
system was a below threshold degenerate optical parametric oscillator. 
The conditioned evolution for this could be solved analytically for 
homodyne detection using a photoreceiver.

The first significant result we found is that for realistic detector 
parameters and realistic systems 
there is a large difference between standard (idealized) 
quantum trajectories and realistic quantum trajectories. Even 
with inefficiencies and dead time included in the standard trajectories, 
the differences are marked. This is due to the effect of finite 
detector response times. For example, for the strongly driven 
two-level system, the average conditional purity under realistic 
direct detection was scarcely better than with no detection at all.

The second significant result we found was the amount of difference 
the detection scheme makes, even using the same detector. With the 
APD, the adaptive scheme gave (for most parameter regimes) 
far higher purity than direct detection. For the PR, homodyne $x$ 
detection (with a local oscillator in quadrature with the system driving 
field) gave a similarly better result than homodyne $y$ detection. 
With a perfect detector the choice of detection scheme of course 
makes no difference as the conditional purity would always be unity.

A final significant result we found was that for homodyne detection 
the PR bandwidth $\gamma$ is not the relevant parameter 
for determining the rates of system evolution that can be tracked. 
Instead, the effective bandwidth is $B = \gamma/\sqrt{N}$. Here $N$ 
is the ratio of electronic noise to vacuum noise, which has been 
assumed small (as required in quantum optics 
experiments). We verified this result numerically for the two-level 
system and analytically for the parametric system. We would expect 
this result to be generalizable to other sorts of detection scheme, 
as long as they involve filtering and the addition of white noise. 
Mesoscopic electronics is an obvious example, as discussed in the 
preceding paper. 

All of these results are relevant to the area of quantum control. As 
explained in the introduction, since a conditional state is a 
representation of the observer's knowledge about a system, it is
by definition  the 
optimal basis for controlling that system. For this to work, the 
observer must have an accurate model for the relation between the 
available information (the detector output) and the quantum system. 
That is precisely what a realistic quantum trajectory is. The 
large differences between standard and realistic quantum trajectories 
noted above means that the former would be a poor approximation to the 
latter in a control system. That is, a realistic quantum trajectory 
theory {\em is} necessary. The differential performance of different 
measurement schemes may also be significant, if one can choose the 
measurement scheme to be used in the control loop.

Finally, another area where 
realistic quantum trajectories would probably 
be needed is in the estimation of 
dynamical parameters for open quantum system from monitoring the 
outputs of such systems. This has been investigated for perfect 
detection in Refs.~\cite{Mab96,GamWis01}. The loss of information 
resulting from realistic detector imperfections will be the subject of 
future work.


\begin{references}
%39
\bibitem{WarWis02a}
P. Warszawski and H.M. Wiseman, preceding paper.
%40
\bibitem{WarWisMab01}
P. Warszawski, H. M. Wiseman and H. Mabuchi,
Phys. Rev. A {\bf 65}, 023802 (2002).
%4
\bibitem{DalCasMol92}
J. Dalibard, Y. Castin and K. M\o lmer,
Phys. Rev. Lett. {\bf 68}, 580 (1992).
%7
\bibitem{WisMil93c}
H. M. Wiseman and G. J. Milburn,
Phys. Rev. A {\bf 47}, 1652  (1993).
%21
\bibitem{WisToo99}
H. M. Wiseman and G. E. Toombes,
Phys. Rev. A {\bf 60}, 2474 (1999).
%22
\bibitem{wtime}
H. J. Carmichael, S. Singh, R. Vyas and P. R. Rice,
Phys. Rev. A, {\bf 39}, 1200 (1989).
%23
\bibitem{Wis96a}
H. M. Wiseman,
Quantum Semiclass. Opt {\bf 8}, 205 (1996).
%25
\bibitem{Dohe99}
A. C. Doherty, S. M. Tan, A. S. Parkins, and
D. F. Walls,
Phys. Rev. A {\bf 60}, 2380 (1999).
%26
\bibitem{BruFinMer02}
T. A. Brun, J. Finkelstein, and N. D. Mermin,
Phys. Rev. A. {\bf 65}, 032315 (2002).
%27
\bibitem{CavesProb}
C. M. Caves, C. A. Fuchs and R. Shack,
Phys. Rev. A. {\bf 65}, 022305 (2002).
%14
\bibitem{Rice88}
P. R. Rice and H. J. Carmichael, IEEE J. Quantum Elect. {\bf 24}, 1351 (1988);
Q. A. Turchette, R. J. Thompson, and H. J. Kimble, Appl. Phys. B {\bf 60}, S1
(1995).
%28
\bibitem{MabPriv}
H. Mabuchi,
California Institute of Technology,
private communication.
%36
\bibitem{newFoc}
www.newfocus.com
%29
\bibitem{GarNoise}
C. W. Gardiner,
{\em Quantum Noise}
(Spring\-er-Verlag, Berlin, 1991).
%30
\bibitem{Jac93}
O. L. R. Jacobs, {\it Introduction to Control Theory},
(Oxford University Press, Oxford, 1993);
P. Whittle, {\it Optimal Control},
(John Wiley \& Sons, Chichester, 1996).
%31
\bibitem{KMan}
F. S. Schweppe,
{\em Uncertain Dynamical Systems}
(Englewood Cliffs, NJ: Prentice-Hall).
%32
\bibitem{HowVac}
H. M. Wiseman and J. A. Vaccaro,
Phys. Lett. A {\bf 250}, 241 (1998).
%33
\bibitem{QOTool}
S. Tan,
{\em Quantum Optics Toolbox for Matlab},
Version 0.10 11-Jan-1999 (University of Auckland).
%34
\bibitem{EulerCarmichael}
H. Carmichael,
University of Oregon,
private communication.
%18
\bibitem{Gar85}
C. W. Gardiner,
{\em Handbook of Stochastic Methods}
(Spring\-er, Berlin, 1985).
%35
\bibitem{Cresser}
J. D. Cresser,
``Ergodicity of Quantum Trajectory Detection Records'' in {\em 
Directions in Quantum Optics} eds. H. J. Carmichael, R. J. Glauber and
M. O. Scully (Springer, Berlin, 2001).
%38
\bibitem{Mab96}
H. Mabuchi,\\
Quan. Semiclass. Opt. {\bf 8}, 1103 (1996).
%39
\bibitem{GamWis01}
J. Gambetta and H.M. Wiseman, 
Phys. Rev. A {\bf 64}, 042105 (2001).

\end{references}
\end{document}